\documentclass[journal]{IEEEtran}
\IEEEoverridecommandlockouts
\usepackage{cite}

\usepackage{amsmath,amssymb,amsfonts}
\usepackage{algorithmic}
\usepackage{graphicx}
\usepackage{textcomp}
\usepackage{xcolor}
\usepackage[ruled,linesnumbered]{algorithm2e} 
\usepackage[nolist]{acronym}
\usepackage{soul}
\usepackage{marginnote}
\usepackage{url}
\usepackage[inline]{enumitem}
\usepackage{tabularx}
\usepackage{amsthm}
\usepackage{tikz}
\usepackage[utf8]{inputenc}
\usepackage[T1]{fontenc}
\usepackage{tikz}

\usepackage{hyperref}
\hypersetup{
    colorlinks=true,
    linkcolor=blue,
    filecolor=magenta,      
    urlcolor=cyan,
    pdftitle={Overleaf Example},
    pdfpagemode=FullScreen,
    }
\newcommand{\myssec}[2]{%
  \subsection{#1}%
  \label{sec:#2}%
}
\newcommand{\rsec}[1]{%
  Sec.~\ref{sec:#1}%
}

\newcommand{\myfigeps}[3][width=\columnwidth]{%
\begin{figure}[htbp!]%
\centering%
\includegraphics[#1]{figures/#2}%
\vspace{-1em}%
\caption{#3}%
\label{fig:#2}%
\end{figure}%
}

\newcommand{\myfigfulleps}[3][width=\textwidth]{%
\begin{figure*}[tb]%
\centering%
\includegraphics[#1]{figures/#2}%
\caption{#3}%
\label{fig:#2}%
\end{figure*}%
}

\newcommand{\rfig}[1]{Fig.~\ref{fig:#1}}

\newcommand{\req}[1]{Eq.~(\ref{eq:#1})}

\newenvironment{myinlinelist}%
{%
\begin{enumerate*}[label=(\roman*)]%
}%
{%
\end{enumerate*}%
}

\newenvironment{myitemlist}%
{%
\begin{itemize}[parsep=0em,leftmargin=*,label={--}]%
}%
{%
\end{itemize}%
}

{%
\begin{enumerate}[parsep=0em,leftmargin=*,label=\arabic*.]%
}%
{%
\end{enumerate}%
}

\newcommand{\keypoint}[1]{\noindent\textit{\underline{Key point:} #1}}
\newcommand{\soadiff}[1]{\noindent\textit{\underline{Key difference:} #1}}

\newcommand{\myurl}[1]{\href{#1}{\url{#1}}}
\newcommand{\myurlfoot}[1]{\footnote{\href{#1}{\url{#1}}}}

\newcommand{\optional}[1]{%
  {\color{red}#1%
  }%
}
\renewcommand{\optional}[1]{#1}

\begin{document}

\title{
  Qkd@Edge: Online Admission Control of Edge Applications with QKD-secured Communications
}

\author{
\IEEEauthorblockN{Claudio Cicconetti}
\IEEEauthorblockA{\textit{IIT, CNR} --
Pisa, Italy \\
c.cicconetti@iit.cnr.it}
\and
\IEEEauthorblockN{Marco Conti}
\IEEEauthorblockA{\textit{IIT, CNR} --
Pisa, Italy \\
m.conti@iit.cnr.it}
\and
\IEEEauthorblockN{Andrea Passarella}
\IEEEauthorblockA{\textit{IIT, CNR} --
Pisa, Italy \\
a.passarella@iit.cnr.it}
}

\author{Claudio~Cicconetti,
        Marco~Conti,
        and Andrea~Passarella%
\IEEEcompsocitemizethanks{\IEEEcompsocthanksitem All the authors are with the Institute of Informatics and Telematics (IIT) of the National Research Council (CNR), Pisa, Italy.}%
}

\IEEEtitleabstractindextext{%
\begin{abstract}
  Quantum Key Distribution (QKD) enables secure communications via the exchange of cryptographic keys exploiting the properties of quantum mechanics.
Nowadays the related technology is mature enough for production systems, thus field deployments of QKD networks are expected to appear in the near future, starting from local/metropolitan settings, where edge computing is already a thriving reality.
In this paper, we investigate the interplay of resource allocation in the QKD network vs.\ edge nodes, which creates unique research challenges.
After modeling mathematically the problem, we propose practical online policies for admitting edge application requests, which also select the edge node for processing and the path in the QKD network.
Our simulation results provide initial insights into this emerging topic and lead the way to upcoming studies on the subject.

\end{abstract}

}

\maketitle

\begin{tikzpicture}[remember picture,overlay]
    \node[anchor=south,yshift=8pt] at (current page.south) {\fbox{\parbox{\dimexpr\textwidth-\fboxsep-\fboxrule\relax}{
      \footnotesize{
      \textcopyright 2023 IEEE.  Personal use of this material is permitted.  Permission from IEEE must be obtained for all other uses, in any current or future media, including reprinting/republishing this material for advertising or promotional purposes, creating new collective works, for resale or redistribution to servers or lists, or reuse of any copyrighted component of this work in other works.
      }
    }}};
\end{tikzpicture}

\IEEEdisplaynontitleabstractindextext

\IEEEpeerreviewmaketitle


%
  \section{Introduction}%
  \label{sec:introduction}%
  Quantum technologies are evolving at a fast pace, with Google recently announcing the achievement of the second of their six milestones towards full commercial exploitation of quantum computing~\cite{acharya_suppressing_2023}, with its immense speed-up of computation time for some specific problems today practically unattainable with classical computers.
On the other hand, the Quantum Internet~\cite{gyongyosi_advances_2022} is slowly taking shape thanks to advancements in quantum communications and ancillary technologies, such as quantum memories.
In the long-term, the Quantum Internet will enable interactions of remote quantum computers, which will unlock a further computation time speed-up, thanks to parallelization and pooling of distant quantum computing resources.
Unfortunately, this scenario requires end-to-end entanglement of quantum bits (qubits) on the remote systems and quantum error correction at intermediate hops: in the scientific community, it is generally agreed that such ``amenities'' will not appear any time soon.

But there is an application of quantum communications that can be used in the short term, since it only relies on a small and technologically-mature subset of the features required by the Quantum Internet: \ac{QKD}~\cite{cao_evolution_2022}.
The latter consists in the secure exchange of cryptographic material that can be used as symmetric keys to ensure private communications between two hosts or applications.
\ac{QKD} relies on the no-cloning property of qubits, which makes the key exchange \textit{unconditionally secure}, and, more importantly, feasible with \textit{commercial} networking equipment.
The most common medium used for QKD is optical fibers, which makes it possible to reuse the existing infrastructure deployed for classical networks.
For instance, in~\cite{bacco_field_2019} the authors demonstrated QKD, with an exchange rate of secret material at 3.4~kb/s, in the metropolitan area of Florence (Italy), with simulations showing scalability up to 175~km when using ultra-low loss single mode fibers.
Conveniently, quantum and classical communications can even share the same cable, though with some mutual performance degradation~\cite{berrevoets_deployed_2022}.
\optional{Significant public and private investments are being done all over the world to deploy quantum communication infrastructures and enable QKD in the near-term (and lay the foundations for the Quantum Internet in the long-term), e.g., the EuroQCI joint initiative between the European Commission and the European Space Agency.}

\myfigeps{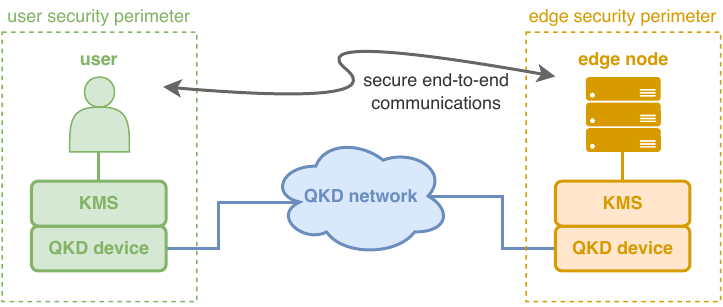}{Basic illustration of QKD-enabled secure communications in an edge computing environment. The user and edge applications obtain symmetric keys, periodically refreshed, by a \ac{KMS} within their  respective security perimeters, which also include the devices producing such keys via QKD, possibly through a QKD network.}

To have practical impact on the market, QKD will have to be integrated within the existing communications and computing infrastructures.
We believe that edge computing~\cite{9519636} provides the perfect environment for early deployments because
\begin{myinlinelist}
    \item many applications, especially in the \ac{IoT} domain, have strict privacy requirements; and
    \item since quantum network links, with optical fibers, are relatively short (realistically within 20-30~km), most of the infrastructures under construction lie within a metropolitan area, though with long-term plans to join these ``islands'' via free-space satellite links~\cite{yin_satellite-based_2017}.
\end{myinlinelist}
At a high level, our reference architecture is shown in \rfig{qkdedge-intro}: the applications on the user vs.\ edge domains establish a secure communication channel through the use of key material provided by the respective \acfp{KMS}, each fed by QKD devices in communication through a QKD network (more details on this in \rsec{basics}).

\noindent\textbf{Contributions:}
In this paper we focus on the following problem, which, to the best of our knowledge, is still uninvestigated.
We envision a processing infrastructure at the edge that provides clients with the opportunity to offload (part of) their computation through the execution of edge applications; furthermore, we assume that there is a QKD network that enables a secure exchange of user data between the clients and edge servers.
When a new edge application requests to enter the system, an admission control procedure is performed and, if admitted, classical/QKD network and processing resources are provisioned accordingly.
We define in a formal manner such problem under idealized conditions, then investigate the options for a practical realization, which leads to the definition of three online policies evaluated using simulations in an unstructured topology with a trace-driven workload composition.


\noindent\textbf{Structure:}
In \rsec{basics} we illustrate the basics of QKD protocols and networks.
The related work is surveyed in \rsec{soa}.
QKD/edge computing admisson control is then discussed in \rsec{model} and evaluated in \rsec{eval}.
\rsec{conclusions} concludes the paper.%

  \section{QKD Basics}%
  \label{sec:basics}%
  The section briefly introduces some basic concepts about \ac{QKD} that are needed to fully appreciate the rest of the paper and it can be skipped by the reader who is already familiar with the topic.
Readers interested in a broader/deeper introduction to QKD are invited to read the survey in~\cite{cao_evolution_2022}.

Quantum information technologies adopt the concept of quantum bits (or qubits), which represent the state of a quantum system $\psi = \alpha |0\rangle + \beta |1\rangle$, which can be in a superposition of the two states $|0\rangle$ and $|1\rangle$ as given by the complex weights $\alpha,\beta$.
Qubits can be realized in static semiconducting circuits, most often used for computations in quantum computers, or carried by photons to implement quantum communications, which is the focus of this paper.
Qubits cannot be copied due to fundamental laws of physics (\textit{no-cloning} theorem): while this can be frustratingly annoying when manufacturing computers, the property is a blessing for communications because it allows to devise protocols to exchange secret material with \textit{unconditional security}.

The first and most famous of such \ac{QKD} protocols is called BB84~\cite{bennett_quantum_2014} and it allows two parties, notoriously called Alice and Bob, to obtain each a sequence of bits following some steps involving the preparation of qubits in a given initial state by one party, their physical transmission to the other party (e.g., carried by photons via optical fiber\footnote{Due to difficulties in engineering the detection of single photons, today weak laser pulses are often used instead, but this is irrelevant to our purposes.}), and a measurement operation by the latter.
In quantum mechanics, measuring the state of a system produces a result in the form of classical information (bits) but at the same time destroys the quantum state.
For this reason, if an eavesdropper has physical access to the quantum communication channel used by Alice and Bob and attempts to read the qubits exchanged, the intended recipient can detect this operation and take appropriate actions, such as repeating some previous steps or giving up.
BB84 also requires some exchanges via a classical channel, which can be public because the eavesdropper cannot gain any information by listening to it.
Once Alice and Bob obtain a sequence of bits that is known only to them, they can use it to generate a symmetric key to cipher messages exchanged over the classical channel.
The \ac{QKD} protocol can be repeated periodically to refresh the secret key.

\myfigeps{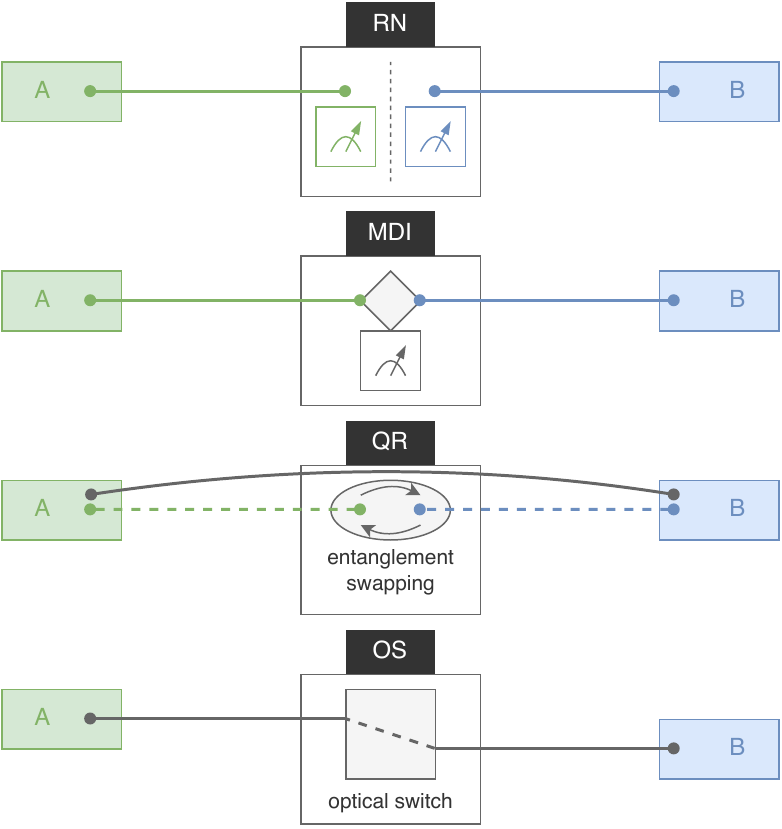}{Different models to extend \ac{QKD} beyond short-range one-hop links to accommodate key exchange between Alice (A) and Bob (B). \textbf{RN}: using \textit{trusted relay nodes}, which terminate the \ac{QKD} protocol at every hop and then perform relaying of secret information via the classical channel (see \rfig{qkd-relay}). \textbf{MDI}: with \textit{heralding stations}, which implement interference between the two qubits from A and B and then a measurement operation. \textbf{QR}: through \textit{quantum repeaters} realizing entanglement swapping, which results in end-to-end entanglement of the qubits at A and B. \textbf{OS}: via \textit{passive optical switches}, creating a single light path between A and B.}

In practice, \ac{QKD} can fully absorb the role played today by public-key cryptosystems, such as RSA\footnote{In this paper we do not compare QKD with classical schemes nor advocate the superiority of one compared to another, which is a matter of (sometimes heated) debate in the research community.}.
However, there are two obstacles to its widespread adoption.
First, qubits are very fragile, because they are (ideally) encoded in \textit{individual} physical particles, such photons.
Therefore, very advanced technologies are needed for quantum communications, which only recently are approaching a maturity level sufficient for the installation in the field and a smooth operation in uncontrolled environments.
All the most mature QKD implementations are based on ``Prepare\&Measurement'' schemes, like BB84 above, which however makes them inherently single-hop: Alice and Bob must be able to exchange qubits without intermediaries or amplifiers, which are prohibited by the no-cloning theorem.
With optical fibers, this limits the maximum communication distance to tens of kilometers.
To achieve a wider coverage, aiming at the Quantum Internet, we need QKD networks, for which different options exist, as illustrated in \rfig{qkd-long-distance}.

\myfigfulleps{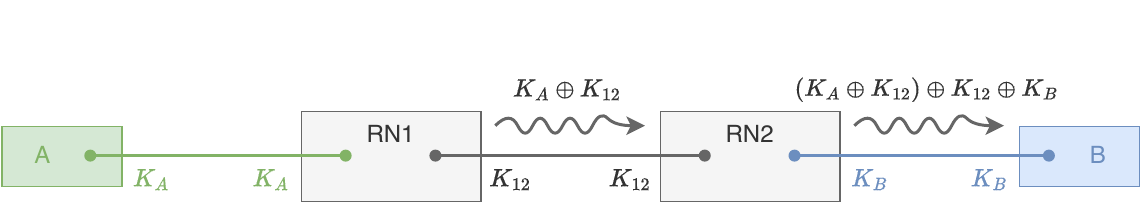}{Example of \ac{QKD} between Alice (A) and Bob (B) relaying via two trusted \acp{RN}. The squiggle arrows ($\rightsquigarrow$) indicate transfer of information in the classical channel and the $\oplus$ sign represents the bit-wise XOR operation. Node B can retrieve the secret key to communicate with Alice $K_A$ by combining together the information from RN2 and its secrete key as $\left(\left(K_A \oplus K_{12}\right) \oplus K_{12} \oplus K_B\right) \oplus K_B = K_A$.}

The easiest way is using intermediate nodes that perform the QKD protocol on each side towards the two end nodes, thus effectively creating isolated segments of secure communications.
These intermediaries are called \textit{\acfp{RN}} and they allow
end-to-end secret keys to be obtained in several ways~\cite{itu-t_quantum_2020}.
The simplest manner, for instance, is to propagate from one end to another on the public classical channel a XOR of the secret key and the cumulative secret keys generated, as shown in the example in \rfig{qkd-relay} with two \acp{RN}.
One disadvantage is that the \acp{RN} need to be \textit{trusted}, otherwise the secrecy of the keys exchanged end-to-end is compromised.
The second option in \rfig{qkd-long-distance} is to use \ac{MDI} \ac{QKD}~\cite{berrevoets_deployed_2022}: the intermediate devices perform interference and measurement operations on the qubits received from A and B, which results in an unconditionally secure exchange of bits between them without the station in the middle acquiring any information on the secret material.
A similar approach consists in the use of \acp{QR}~\cite{rozpedek_near-term_2019}, which perform an operation called entanglement swapping that creates an entanglement between the qubits on the end nodes, such that multiple physical hops can be combined into a single logical hop.
Despite some encouraging initial results on laboratories and research infrastructures, solutions based on both MDI and QR today are far less robust and mature than those relying on \acp{RN}.
Finally, we mention that passive optical switches (bottom part of \rfig{qkd-long-distance}) can also be used to create physical communication paths in an optical QKD network~\cite{azuma_all-photonic_2015}, even though this option does not extend the range of installations but only improves their flexibility.

\keypoint{While QKD devices are ready to production deployments (manufacturers include Toshiba and ID Quantique), the design and operation of efficient QKD networks is still an open research challenge, with short-term solutions relying on the use of trusted \acp{RN}.}%

  \section{Related Work}%
  \label{sec:soa}%
  In this section we provide a short survey on the scientific literature relevant to our work, which covers QKD networks and edge computing.
The latter has been very popular in the last decade and there are countless prior works, thus it would be daunting as much as pointless to summarize them all here.
Instead, we limit ourselves to referring the interested reader to a recent survey paper~\cite{9519636} and highlighting the following:

\keypoint{While some of the studies in the literature of edge computing can be considered influential for the topic of this paper, to the best of our knowledge there are no previous works that consider the allocation of processing resources in an edge infrastructure that includes QKD, which has distinguishing features compared to classical networks. These include: limited resources (as the technology is in its infancy), unstructured topology (because of the maximum length of a single hop), constant rates (driven by the applications' security requirements).}

In the remainder of this section we address the more relevant literature on QKD, which is, in general, by far more modest.
In~\cite{mehic_novel_2020} the authors noted some similarities between QKD networks with \acp{RN} and \acp{MANET}: limited one-hop distance, capacity inversely proportional to distance, hop-by-hop routing, links unreliable (due to lack of key material and possibly to the public channel becoming congested~\cite{mehic_analysis_2017}).
Thus, they proposed a distributed geographic reactive routing model to minimize the number of routing packets and achieve high-level scalability, which was compared to OSPFv2 with simulations.
\soadiff{In their work it is assumed that resource allocation in the QKD network is disjoint from the upper layers, which we also consider as an option in \rsec{model:online} below. However, in this paper we do not delve into the details of how to realize routing as in the work~\cite{mehic_novel_2020}, which can thus be considered complementary to ours.}

A layered architecture of a complete scalable QKD system is illustrated in~\cite{tysowski_engineering_2017}, where the authors propose to assign routes in the QKD network by solving a maximum concurrent multi-commodity flow problem, which maximizes the throughput assigned to selected pairs of users.
\soadiff{The approach proposed is elegant and relatively efficient, because well-known good approximations exist for that problem, but ours is different because we have fixed demands, coming from the joint allocation of edge computing applications.}

In~\cite{cao_kaas_2019} a key-as-a-service framework is illustrated, which provides keys in a reactive virtualization-based manner, while a secret-key-aware routing method is investigated in~\cite{yang_quantum_2018}, which proposes a dynamic forecast mechanism for the residual local key availability at each link. 
\soadiff{Both these studies are complementary to ours, where the availability of QKD (and computing) resources is checked before admission.
A comparison between on-demand and reservation-based schemes is left for future work.}

Finally, we mention that a complete stack for the integration of QKD in a cloud-native environment was proposed in~\cite{pedone_toward_2021}, where the authors focused on the interfaces between the main components, such as QKD devices and the \ac{KMS}.
This work confirms the interest of the research community on the use and optimization of QKD systems as an integral part of upcoming telecommunicatons/computing infrastructures, as also corroborated by the authors of~\cite{lopez_applying_2019}, who presented realistic use cases from a telco provider's perspective.




%

%
  \section{QKD/Edge Admission Control}%
  \label{sec:model}%
  In this section we illustrate the system model and define a mathematical formulation for joint resource allocation (\rsec{model:model}), followed by a discussion (\rsec{model:discussion}), and the proposal of three online admission control policies (\rsec{model:online}) evaluated in the next section.

\myssec{System model and mathematical formulation}{model:model}

As introduced in \rsec{introduction}, we envision an edge computing infrastructure where the clients offload their computation needs (totally or partially) to a number of edge nodes with computation power operated by a \textit{service provider}.
The network connecting the clients to the edge nodes, operated by the \textit{network operator}, is mixed classical/QKD.
In general, some (fiber optic) links can be used for both classical and quantum communications, while others can be only classical (e.g., terrestrial wireless links) or QKD (e.g., satellite links).
Computation offloading is done through the use of \textit{microservices}, which is the dominating paradigm in today's cloud-native architectures: applications are assigned virtual environments (most often containers) where edge-side application images are executed.
Edge nodes have a finite capacity to host microservices, depending on the physical availability and the applications' demands, in terms of CPU/memory/storage.

\myfigeps{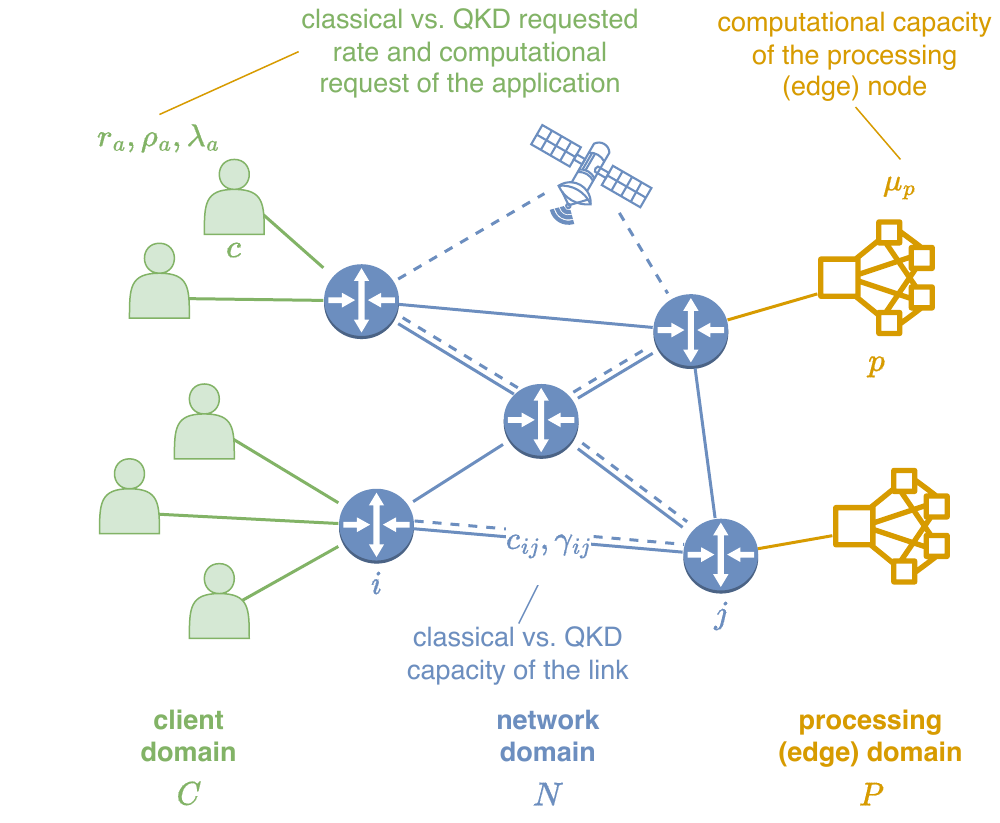}{System model illustrating the client, network, and edge domains and the corresponding notation. The dashed lines represent QKD links, while solid ones are classical links.}

The system model is illustrated in \rfig{qkdedge-model}, showing the three domains: client, network (classical and QKD), and edge.
For convenience of mathematical formulation, we model the system as a graph $G(V,E)$, where $V=C \cup N \cup P$ are the clients ($C$), network devices ($N$), and edge nodes ($P$), also called \textit{processing nodes} in the following. There is an edge $(i,j) \in E$ iff two components $i$ and $j$ can communicate directly.
$A$ is the set of applications in the system, where each application $a \in A$ is hosted by a client $c_a \in C$.

\underline{QKD/Edge Fractional Allocation Problem (QEFAP)}.
Input:

\begin{myitemlist}
    \item For each application $a \in A$: $r_a$ is the traffic rate for classical communications, in b/s; $\rho_a$ is the key rate for QKD communications, in b/s; $\lambda_a$ is the amount of computation needed in the unit of time, expressed in the same unit of measurement as the edge node capacity below.
    \item For each link $(i,j) \in E$: $c_{ij}$ is the capacity of the link for classical communications; $\gamma_{ij}$ is the capacity of the link for the local generation of secret material using QKD.
    For a given link, it can be $c_{ij}=0$ if it is not used for classical communications (dashed line only in \rfig{qkdedge-model}) or $\gamma_{ij}=0$ if the two nodes do not share a QKD communication link (solid line only in \rfig{qkdedge-model}).
    \item For each processing node $p \in P$: $\mu_p$ is the total computational capacity that can be offered to the clients, in the same unit of measurement as $\lambda_a$.
    \item There is a profit $\pi_{ap} \geq 0$ to be gained from allocating app $a$ on edge node $p$.
\end{myitemlist}

\noindent The problem is to find:

\begin{eqnarray}\label{eq:obj}
    \forall a \in A, \forall p \in P, \forall (i,j) \in E: \\
    x_{ap}, f_a(i,j), \phi_a(i,j) & s.t. & \max \sum_{a \in A} \sum_{p \in P} x_{ap} \pi_{ap}, \nonumber
\end{eqnarray}
where $x_{ap} \in [0,\lambda_a]$ is the fraction of computational capacity on edge node $p$ allocated to app $a$;
$f_a(i,j) \leq c_{ij}$ is the classical data traffic flow of app $a$ traversing link $(i,j)$, and
$\phi_a(i,j) \leq \gamma_{ij}$ is the QKD data traffic flow of app $a$ traversing link $(i,j)$, subject to the following constraints.

\noindent The computation capacity of edge nodes is limited:
\begin{equation}
    \forall p \in P: \sum_{a \in A} x_{ap} \leq \mu_{p}.
\end{equation}

\noindent The classical/QKD link capacity is limited:
\begin{eqnarray}
    \forall (i,j) \in E: & \sum_{a \in A} f_a(i,j) \leq c_{ij} \\
    & \sum_{a \in A} \phi_a(i,j) \leq \gamma_{ij}.
\end{eqnarray}

\noindent Classical and QKD flows in transit nodes are conserved:
\begin{align}
    \forall a \in A,\forall n \in N : & \sum_{k \in V} f_a(n,k) - \sum_{k \in V} f_a(k,n) = 0 \\
    & \sum_{k \in V} \phi_a(n,k) - \sum_{k \in V} \phi_a(k,n) = 0.
\end{align}

\noindent Classical and QKD flows at the clients are conserved:
\begin{align}
    \forall a \in A: & \sum_{k \in V} f_a(c_a,k) - \sum_{k \in V} f_a(k,c_a) = r_a \\
    & \sum_{k \in V} \phi_a(c_a,k) - \sum_{k \in V} \phi_a(k,c_a) = \rho_a.
\end{align}

\noindent Classical and QKD flows at the edge nodes are conserved:
\begin{align}
    \forall a \in A, & \forall p \in P: \nonumber \\ 
    & \sum_{k \in V} f_a(k,p) - \sum_{k \in V} f_a(p,k) = r_a \cdot x_{ap}/\lambda_a \\
    & \sum_{k \in V} \phi_a(k,p) - \sum_{k \in V} \phi_a(p,k) = \rho_a \cdot x_{ap}/\lambda_a.
\end{align}

\myssec{Discussion}{model:discussion}

The QEFAP can be solved efficiently using \ac{LP} tools, e.g., the network simplex method.
However, in this formulation, a single application can be allocated to multiple edge nodes, and each edge node can be reached through multipath communications in the classical and QKD networks.
In practice, this can be very difficult or impossible to achieve.
First, spreading a single app across different edge nodes requires the multiple edge applications to maintain a \textit{distributed state}, with the complexity to guarantee consistency growing more than linearly with the number of parties.
Furthermore, the app would have to load balance towards the different targets based on the $x_{ap}$ values\footnote{A possible solution to this problem is adopting a serverless paradigm, where function invocations are stateless: such an approach would not require complex state synchronization on the edge applications and the load balancing can be moved to a dedicated component in common for all the client apps, as done in \cite{cicconetti_toward_2020}.
However, not all the applications can be efficiently realized with this pattern.
We plan to investigate in a future work this opportunity.}.
Second, \textit{multipath communications} are only possible if there is a fine control on the network infrastructure, which is not always the case due to both administrative constraints (the network operator might not be willing to share with its customers a detailed view and fine control of its network) and technical considerations (the complexity and overhead of handling per-flow state in a large network can be overwhelming).

Based on these considerations, we believe that the following \ac{ILP} formulation would be more practical:

\underline{QKD/Edge Integer Allocation Problem (QEIAP)}: Same as QEFAP with the additional constraints that an application $a$ can be assigned to at most one processing node $p$ and the entire flow $c_a \rightarrow p$ must follow the same path in the classical and QKD network, respectively.

QEIAP can be seen as an extension of the \textit{Generalized Assignment Problem}: in addition to finding the best assignment of tasks (clients) to workers (edge nodes) that maximizes the profit, a solution is only feasible if it possible to find paths in the classical and QKD networks that satisfy the capacity constraints.
The problem is NP-hard with no well-known general-purpose efficient heuristics.

In any case, both QEFAP and QEIAP are offline problems: they assume that the \textit{entire input} is known at the time when a solution is sought.
However, the real system would be dynamic: new applications will arrive asynchronously with respect to one another, and likewise they will terminate following independent lifecycles driven by the user needs.
This consideration leads us to the final form of our problem:

\underline{QKD/Edge Online Admission Problem (QEOAP)}: At time $t$ of arrival of application $a$, with given characteristics $r_a,\rho_a,\lambda_a$, determine if it is possible to admit it based on the following two conditions:
\begin{myinlinelist}
    \item there is a non-empty set of processing nodes $P'$ with residual computational capacity $\mu_p(t) \geq \lambda_a$;
    \item there is at least a processing node $\bar{p} \in P'$ with a path in the classical (respectively, QKD) network such that all the intermediate links $(i,j)$ have sufficient residual capacity to carry the required traffic, i.e., $c_{ij}(t) \geq r_a$ (respectively, $\gamma_{i,j}(t) \geq \rho_a$).
\end{myinlinelist}

The attentive reader will have noticed that the transformation of the problem from allocation to online admission has led us to drop the profit in \req{obj}.
An objective function can be re-introduced for all the cases where an application $a$ enjoys multiple feasible options, as discussed in the next section.

\myssec{Online policies}{model:online}

We now develop three policies to solve the QEOAP, which will be called \texttt{SPF}, \texttt{BF}, and \texttt{ANY}.
For simplicity of notation, we assume that classical networks are overprovisioned with respect to the applications' demands, thus they are not considered in the policies below.
Furthermore, we consider a practical scenario where the QKD network operator and service provider (as introduced in \rsec{model:model}) are different entities, which cooperate in providing a commercial offer to customers.
Under this assumption, the admission control of new applications can be done by:
\begin{myinlinelist}
    \item the QKD network operator,
    \item the service provider, 
    \item or, a third party.
\end{myinlinelist}
In all cases, this will result in the assignment of the incoming application to an edge node and the selection of the path in the QKD network.
Below we consider the three cases separately.

\underline{QKD network operator:}
From the point of view of the network operator, in the absence of extra information that is not covered by our system model in \rsec{model:model}, a very reasonable choice would be to select the path that minimizes the number of hops, since this consumes the least amount of network resources.
This strategy is advocated generally by the research community for QKD (e.g., in~\cite{cao_cost-efficient_2019}).
The network operator must possess the information on which edge nodes are available to onboard the new application, called $P'(\lambda_a)$, which depends on the instantaneous load of the processing infrastructures and active applications.
The policy \textbf{\texttt{SPF}} selects the path $\bar{\nu}$ as follows:
\begin{align}
\bar{\nu} = \arg\min_{\nu \in \mathcal{P}(c_a, P'(\lambda_a))} \left\{
    |\nu| : \forall (i,j) \in \nu, \gamma_{ij} \geq \rho_a
 \right\},\nonumber
\end{align}
\noindent where $|\nu|$ is the length of the path $\nu$, in number of hops,  and $\mathcal{P}(c_a,P'(\lambda_a))$ is the set of possible paths from the application's client node $c_a$ to any of the processing nodes $p \in P'(\lambda_a)$.
Automatically, the edge node assigned is the last node in $\bar{\nu}$.
The policy can be implemented in polynomial time using constrained shortest path routing techniques from the traffic engineering domain.
We do not elaborate further on this aspect due to limited page budget.

\underline{Service provider:}
In this case, it is up to the QKD network operator to determine the set of processing nodes $P'(\rho_a)$ that are feasible, i.e., which can be reached by the application's client node under the specified key rate requirement, and make this information available to the service provider.
The latter can then choose the ``best'' processing node based on its objective function; if no extra information is available, we propose to use a best-fit approach, inspired from the operations research literature.
In fact, best-fit is a widely used approximation for the bin packing problem, which reduces the fragmentation of resources, therefore it is often adopted in virtual memory management within operating systems.
In our notation, the policy \textbf{\texttt{BF}} selects the processing node $\bar{p}$ as follows:
\begin{align}
\bar{p} = \arg\min_{p \in P'(\rho_a)} \left\{
    \mu_p - \lambda_a : \mu_p \geq \lambda_a
\right\},\nonumber
\end{align}
\noindent that is $\bar{p}$ picks the edge node which leaves the smallest residual processing capacity.
Also this policy can be implemented efficiently through a wise use of data structures (not elaborated here).

\underline{Third party:}
Finally, we consider the case where both the QKD network operator and the service provider make available the set of feasible nodes $P'(\rho_a)$ and $P'(\lambda_a)$, respectively, but the selection is then taken by a third party.
The latter could be, for instance, the application itself or a load balancer.
In the \textbf{\texttt{ANY}} policy, we simply admit the application if $P'(\rho_a) \cap P'(\lambda_a) \neq \emptyset$, in which case a processing node at random is chosen.


All the policies above require an exchange of real-time information about the feasibility of edge nodes depending on the network/load conditions, i.e., $P'(\rho_a)$ and $P'(\lambda_a)$.
It would be interesting to explore the means to achieve this in practice through the use of standard management/control interfaces, such as those defined by ETSI MEC, but we consider the issue beyond the scope of this work.%

  \section{Performance Evaluation}%
  \label{sec:eval}%
  In this section we evaluate the performance of the different policies for online assignment of applications to edge nodes through a QKD network defined in \rsec{model:online}.
We first illustrate the methodology adopted (\rsec{eval:model}), then we introduce two further policies defined for the purpose of establishing a performance baseline (\rsec{eval:policies}), and finally we report and discuss the results found (\rsec{eval:results}).

\myssec{Methodology, assumptions, and tools}{eval:model}

\noindent\textbf{Tool:}
Due to the lack of large-scale QKD networks in production, we evaluated the performance using simulation with QueeR\myurlfoot{https://github.com/ccicconetti/queer} (tag \texttt{v1.6}, experiment \texttt{006}), which we maintain as part of our research activities and was used already for studies on traffic engineering of entanglement-based quantum networks.
The software is open source and shipped with simulation scripts and artifacts for full reproducibility.
\optional{
Even though quantum networking is an emerging field of research, there are already several open-source simulation tools available, each best tailored to a specific type of analysis.
The alternatives include: SimulaQron\myurlfoot{http://www.simulaqron.org/}, which is intended to assist quantum network application development, the QKD Simulator\myurlfoot{https://www.qkdsimulator.com/}, which focuses on the performance of the QKD protocol itself, and add-ons of packet level simulators (QKDNetSim\myurlfoot{https://openqkd.eu/qkd-network-simulator/} for ns-3, QuISP\myurlfoot{https://aqua.sfc.wide.ad.jp/quisp_website/} for Omnet++), which are suited to the evaluation of QKD/classical integrated services.
To the best of our knowledge, none of the other tools available would fit precisely our plan, as in this paper we are interested in the steady-state system performance of a mixed QKD network/edge computing infrastructure.
For this reason, to match the purposes in this research activity, we have extended our software QueeR.}

\noindent\textbf{Methodology:} For the simulations we adopted a Monte Carlo approach: for each scenario we repeatedly select a random topology and a random workload of applications that are assumed to have requested the admission (see below for details); then, we perform the assignment of applications to the edge nodes one after another, to emulate an online procedure where applications enter the system asynchronously.
In the plots in \rsec{eval:results} we report the average values from all such independent replications, whose number (5000) is such that a relatively small variance was measured (not shown in the plots for better readability).

\noindent\textbf{Topology:} The topology is generated using the Waxman model~\cite{waxman_routing_1988}, which is widely adopted in the performance evaluation of quantum networks without a regular structure (e.g.,\cite{mehic_novel_2020}).
Accordingly, in our simulations the nodes are uniformly distributed in a plane with maximum inter-node distance $L=100$~km and the probability of two nodes being connected is $p = \beta \cdot e^{-d/\alpha L}$, where $d$ is the distance (in km) between the nodes, $\alpha$ is a parameter that controls the density of short edges compared to that of longer ones, and $\beta$ is a parameter that drives the edge density.
Unless specified otherwise, we assume $\alpha = \beta = 0.4$, which are typical values used in the literature.
The rate of QKD links is assumed to decrease exponentially with distance, which is reasonable due to attenuation/absorption effects: $\gamma = \gamma^{\max} e^{-d/L}$, where $\gamma^{\max}=30$~kb/s unless specified otherwise.
\optional{Consistently with our illustration of the policies in\rsec{model:online}, and to simplify the analysis while focusing our attention on QKD-related aspects, we assume that the classical network links always have sufficient capacity for the applications, that is $c_{ij} = \infty$.}
The edge nodes are selected randomly from the set of nodes in the topology.

\noindent\textbf{Workload:} For the workload, we adopted a trace-driven approach starting from a dataset of traces obtained in 2020 from a production system and publicly released by Microsoft~\cite{10.1145/3472883.3486974}.
The dataset contains a log of read/write backend activities of user applications, with associated geographical regions.
We transformed the traces to match the input required by our problem as follows.
The probability than an application requests admission is proportional to its lifetime within the trace; each region is mapped randomly to a node in the network; the application load $\lambda_a$ is equal to the event rate in the trace (ranging from 0.008 to a capped value of 1); the QKD capacity requested by the application $\rho_a$ is the common logarithm of the traffic rate (values range from 1~kb/s to 8~kb/s).
The trace transformation tool is publicly available on GitHub\myurlfoot{https://github.com/ccicconetti/support/tree/master/Dataset/002_Lifecycles} (tag \texttt{dataset-002}).

\myssec{Comparison policies}{eval:policies}

To establish a baseline and provide further insights on the algorithms in \rsec{model:online}, we also evaluated the performance with the following comparison policies.
With \texttt{STATIC} we pre-assign each client node $u$ to its closest edge node $v$, i.e., that with the shortest path in number of hops.
All incoming applications associated to $u$ will be assigned to edge node $v$, if there are sufficient residual computation resources, otherwise they are blocked.
This policy corresponds to a zone-based edge assignment, where the network operator and service provider do not exchange real-time data about the utilization of their respective resources.
Also, such policy is most efficient: the shortest-path assignment is done offline, while at run-time the admission control requires a mere table look-up.
On the other hand, with \texttt{RND} we select an edge node at random: if there are sufficient resources \textit{a posteriori} then the application is accepted, otherwise it is blocked.
\optional{An admission control example with all the policies defined in this work is in \rfig{qkdedge-example}.

\myfigeps{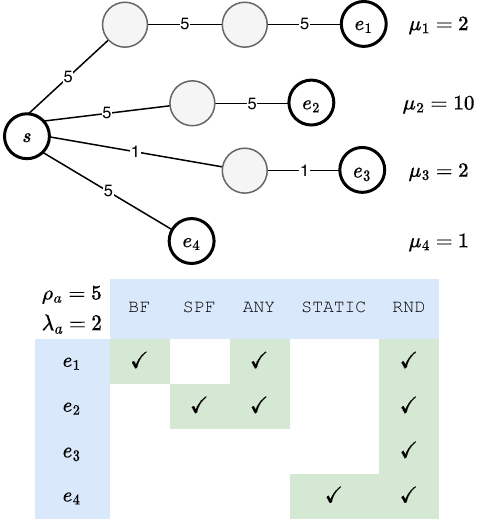}{Example of admission request of application $a$ ($\rho_a = 5$~kb/s, $\lambda_a = 2$) on client node $s$ with the five policies compared in \rsec{eval:results} in a simple QKD network topology with given residual processing availability $\mu_1,\ldots,\mu_4$ on the four edge nodes $e_1,\ldots,e_4$; all the links have QKD capacity equal to 5~kb/s, except those connecting $s$ to $e_3$, for which the capacity is 1~kb/s. Ties are broken randomly with multiple choices. When $a$ is assigned to $e_3$ or $e_4$, this will result in the request being blocked due to insufficient resources, QKD capacity or residual processing availability, respectively.}}

\myssec{Results}{eval:results}

\begin{figure*}[tb]%
    \centering
    \includegraphics[width=0.45\textwidth]{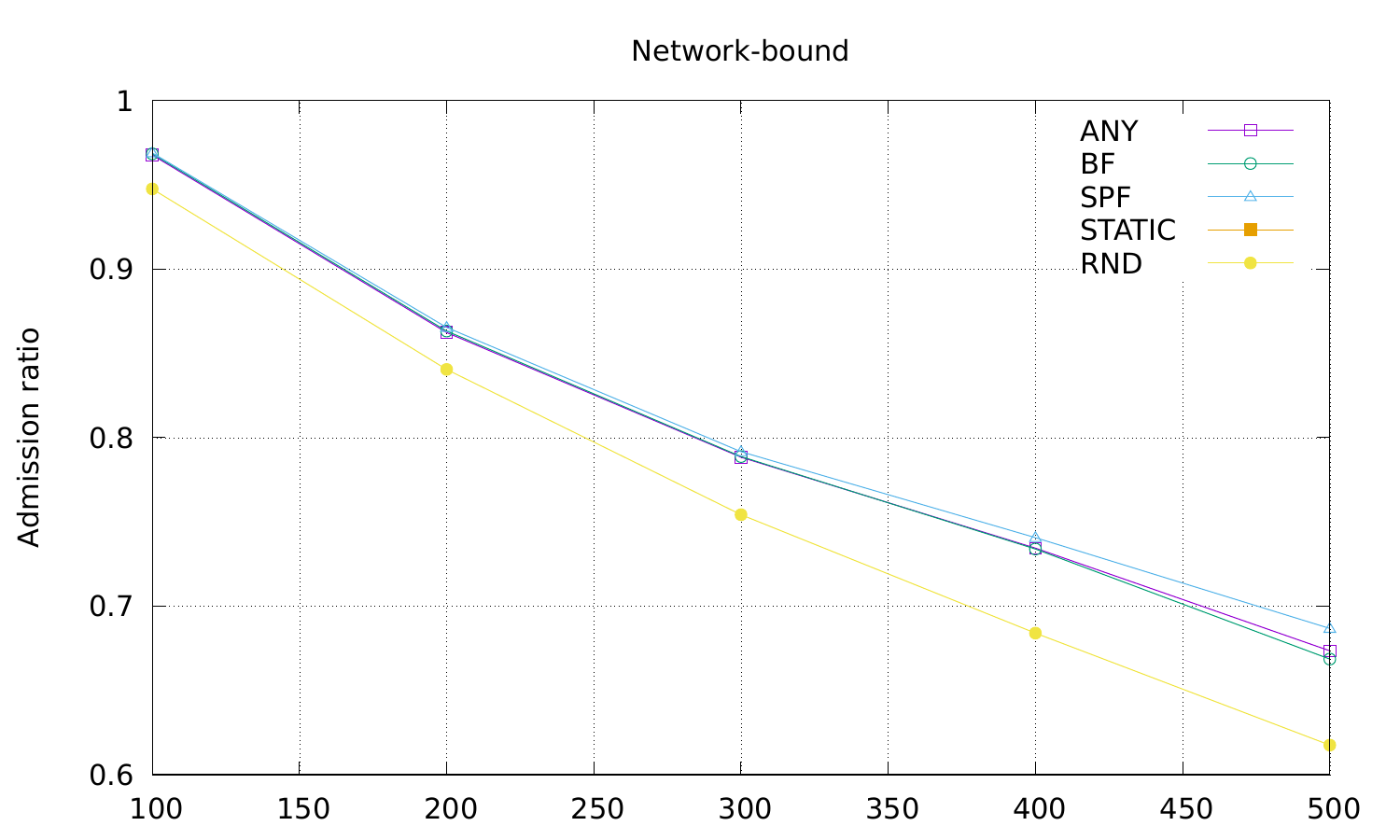}
    \includegraphics[width=0.45\textwidth]{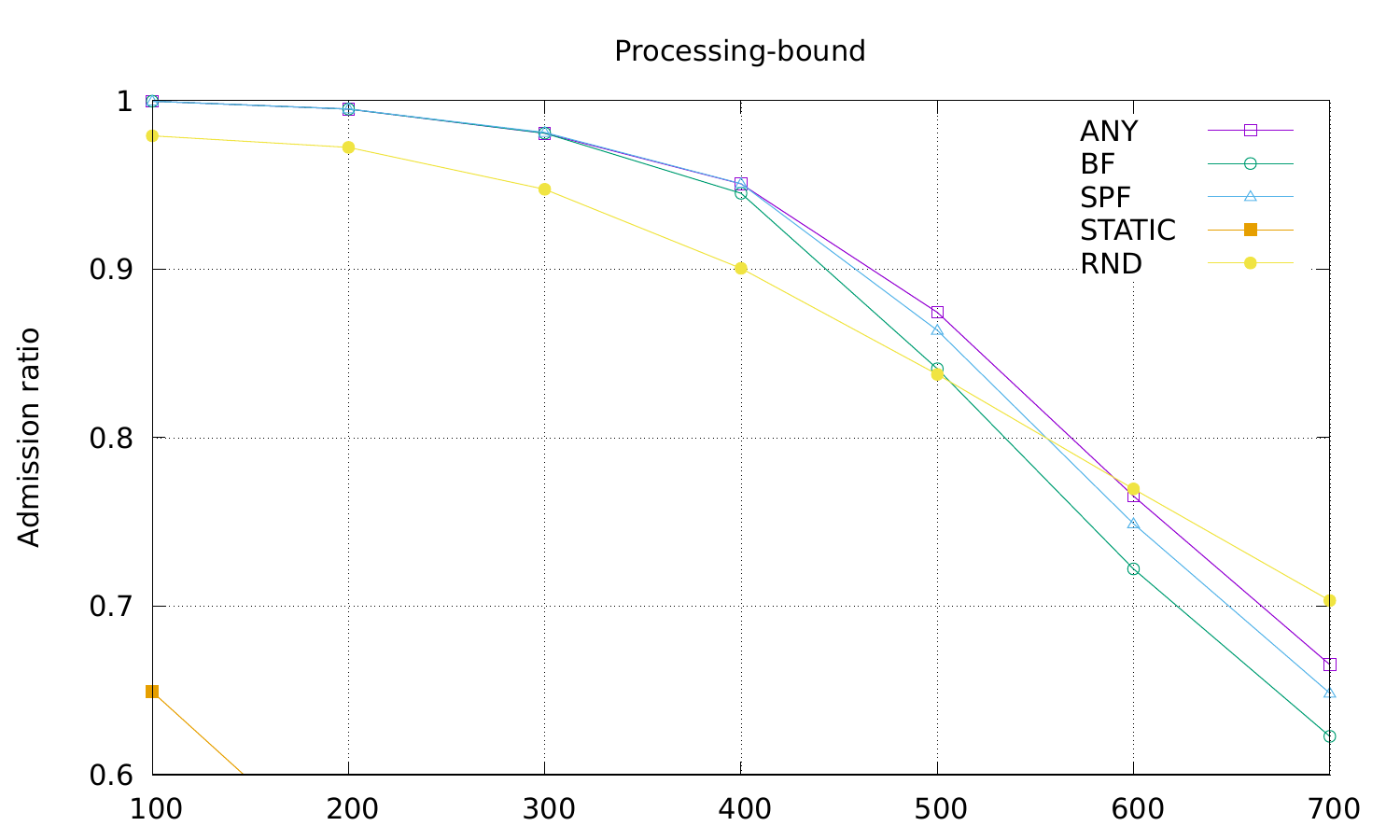}\\
    \includegraphics[width=0.45\textwidth]{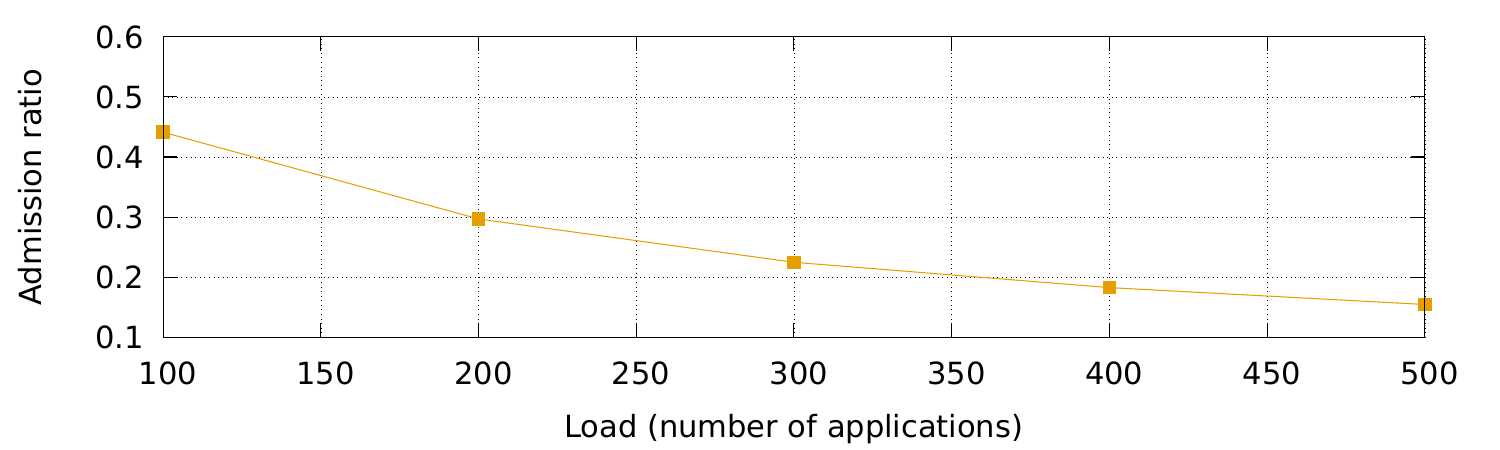}
    \includegraphics[width=0.45\textwidth]{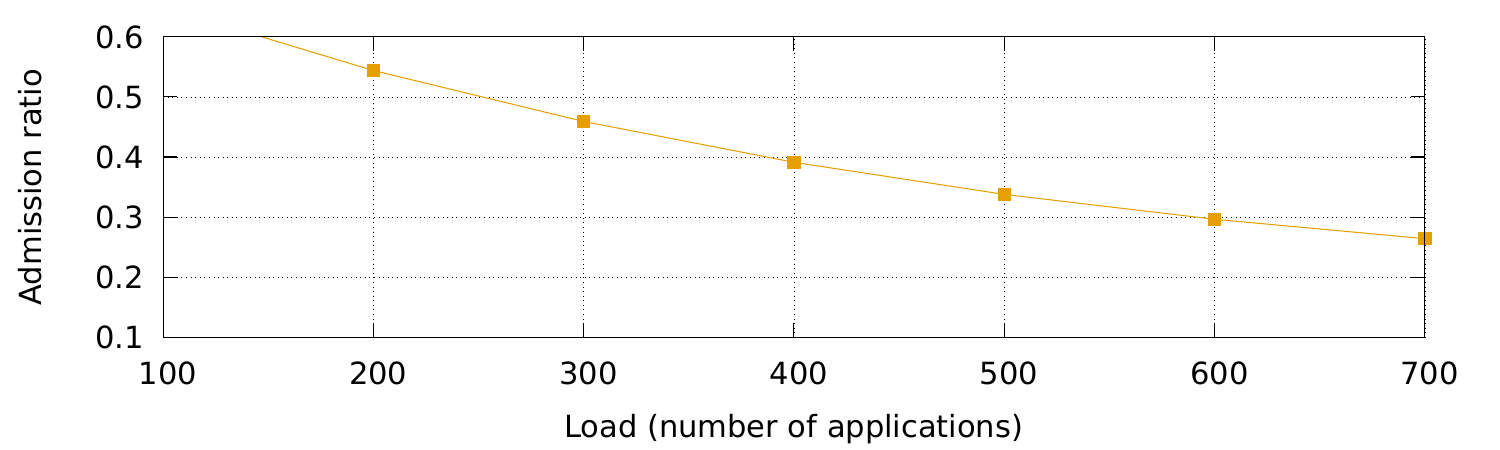}\\
    \caption{Admission ratio with increasing load, in number of applications. Different policies are shown: \texttt{ANY}, \texttt{BF} and \texttt{SPF} are defined in \rsec{model:online}, \texttt{RND} and \texttt{STATIC} in \rsec{eval:policies}. The curve for the latter is shown in a separate plot with different y-range for better readability. The network has 50 nodes (10 are edge nodes, each with capacity $\mu=5$) and it was generated with $L=100$~km, $\alpha=\beta=0.4$. Two values of the maximum QKD link capacity are considered: on the left it is $\gamma^{\max}=30$~kb/s, which means that the QKD network is the choke point, while on the right it is $\gamma^{\max}=100$~kb/s, which shifts the bottleneck to the processing.}
    \label{fig:006-admission}
\end{figure*}

Through initial calibration (results not shown) we have found the range of $\gamma^{\max}$ that determines whether the system performance is limited by the QKD network or edge computing processing capabilities.
We have then run two scenarios with representative values of $\gamma^{\max}$ for the two conditions: 30~kb/s (network-bound) vs.\ 100~kb/s (processing-bound).
In \rfig{006-admission} we show the admission ratio for the two cases while increasing the load, in number of applications that request to be admitted.
First, \texttt{STATIC} performs much worse than all policies, which is explained by the uneven distribution of the clients in the production trace.
\texttt{RND} too has a significantly lower admission ratio, except for very high loads in the processing-bound case.
Bewilderingly, in these conditions it outperforms all the policies (see right plot in \rfig{006-admission} with 700 applications).
This phenomenon is a side effect of the fragmentation of edge node availabilities, which tends to reject applications with a high load request: in fact, up until 500 applications, the admission ratio is low.
However, such an ``early rejection'' is beneficial when there are many applications that are denied admission anyway due to the high load, because it keeps the average load per admitted application lower.
Regarding \texttt{ANY}, \texttt{BF}, and \texttt{SPF}: their behavior is almost overlapping in the network-bound case, while in the processing-bound case and above 600 applications \texttt{ANY} is marginally better than the others.
Interestingly, the network-bound curves decrease almost linearly even at low loads, whereas the processing-bound curves remain close to 1 until the system is overloaded, at which point they plunge.
\keypoint{From the results, it is not possible to identify one policy among \texttt{ANY}, \texttt{BF}, and \texttt{SPF} that clearly outperforms the others in terms of the admisson ratio.
Further experiments, left for future work, are needed to understand if other topologies or workload conditions can lead to different conclusions.}

\myfigeps{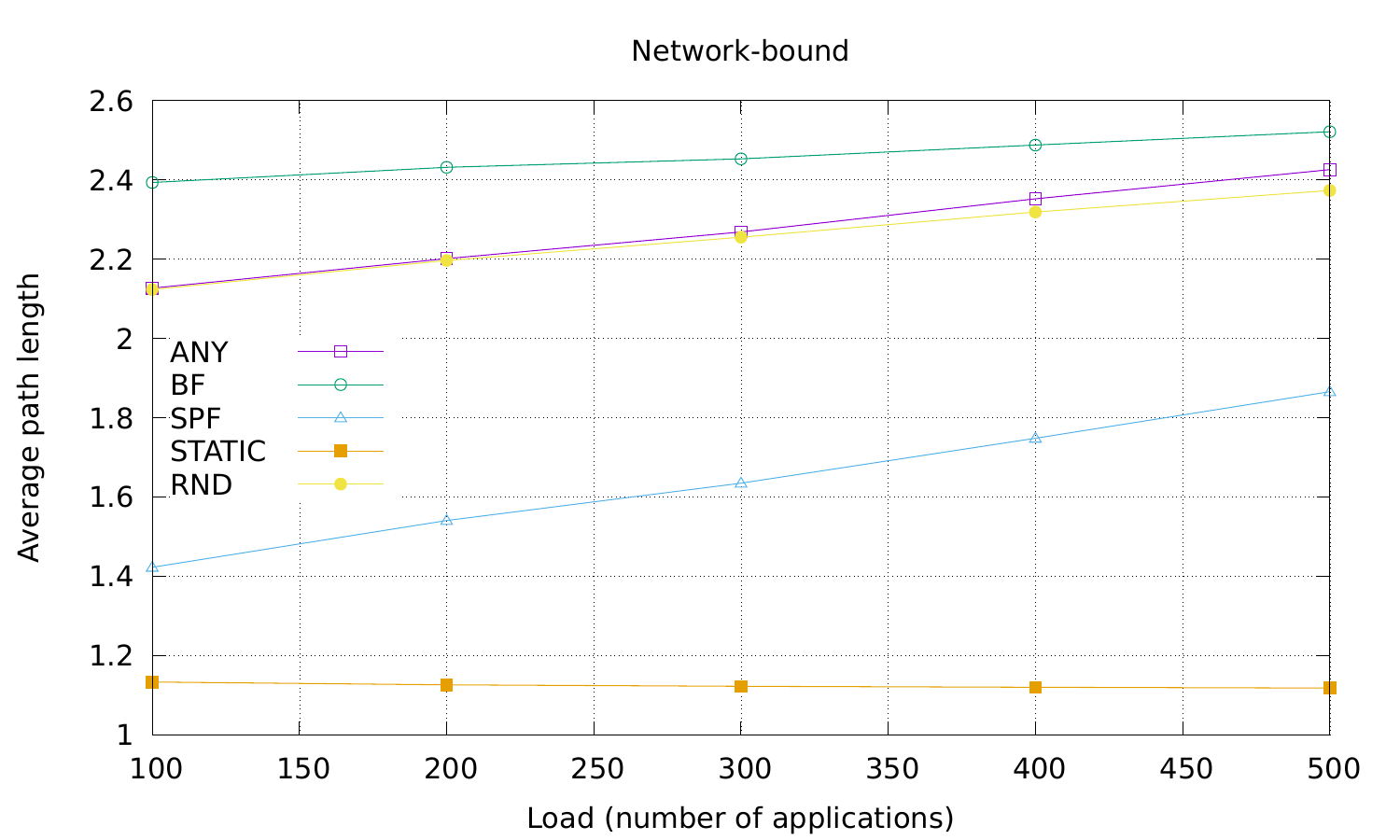}{Average path length with increasing number of applications for the different policies in the network-bound scenario (left case in \rfig{006-admission}).}

\myfigeps{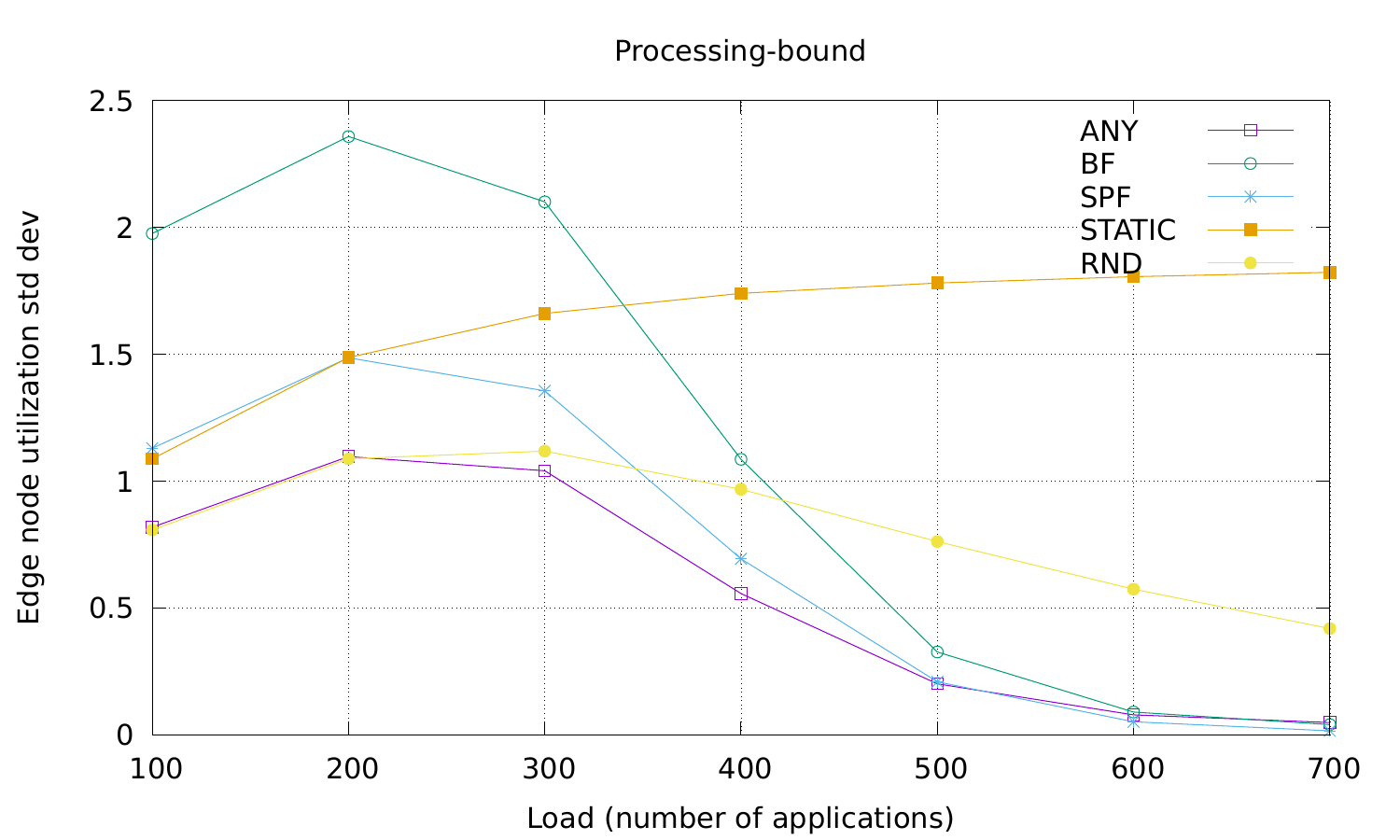}{Standard deviation of the edge node utilization values with increasing number of applications for the different policies in the processing-bound scenario (right case in \rfig{006-admission}).}

Meanwhile, to better understand the results so far, we report in \rfig{006-van-path-length} the average path length of accepted applications for the network-bound case.
As expected, the \texttt{STATIC} policy is independent from the load, while all the others increase.
In particular, \texttt{SPF} lies significantly below all others, but this reflects only slightly on the admission ratio as we have seen.
On the other hand, for the performance-bound case, we show in \rfig{006-val-edge-node-util-stddev} the standard deviation of the residual processing availability on edge nodes, as a measure of how evenly a given policy assigns the applications.
When the load is low, there are significant differences: in particular, \texttt{BF} exhibits a higher variance because it assigns each new application to where it fits ``best''.
However, as the load increases (and the admission ratio drops), the policies in \rsec{model:online} tend to saturate all the edge nodes, thus the variance vanishes.
This is not the case for the comparison policies \texttt{RND} and \texttt{STATIC} which leave a significant portion of resources unused despite the high rejection rate.
\keypoint{\texttt{SPF} and \texttt{BF} exhibit very different behavior, since they aim at minimizing network vs.\ processing resources, and their performance is matched or exceeded by \texttt{ANY}, which combines real-time exchange of information between the network operator and service provider; such policies perform much better than simpler ones that do not require interactions between entities (\texttt{STATIC} and \texttt{RND}), except for some limit cases of minor practical relevance.}

\myfigeps{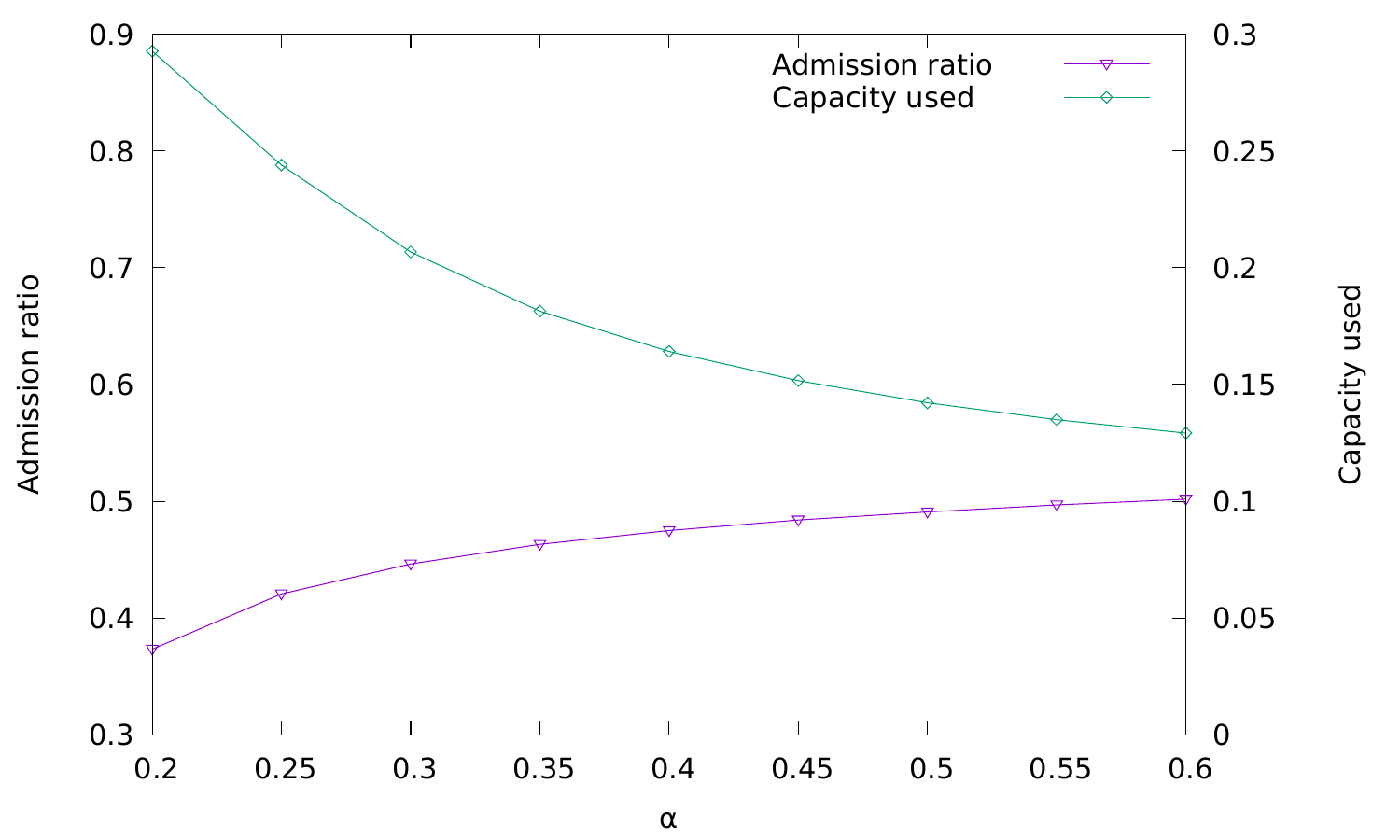}{Admission ratio and capacity used with increasing $\alpha$ values with \texttt{SPF} policy, 50 nodes, and 300 applications.}

\myfigeps{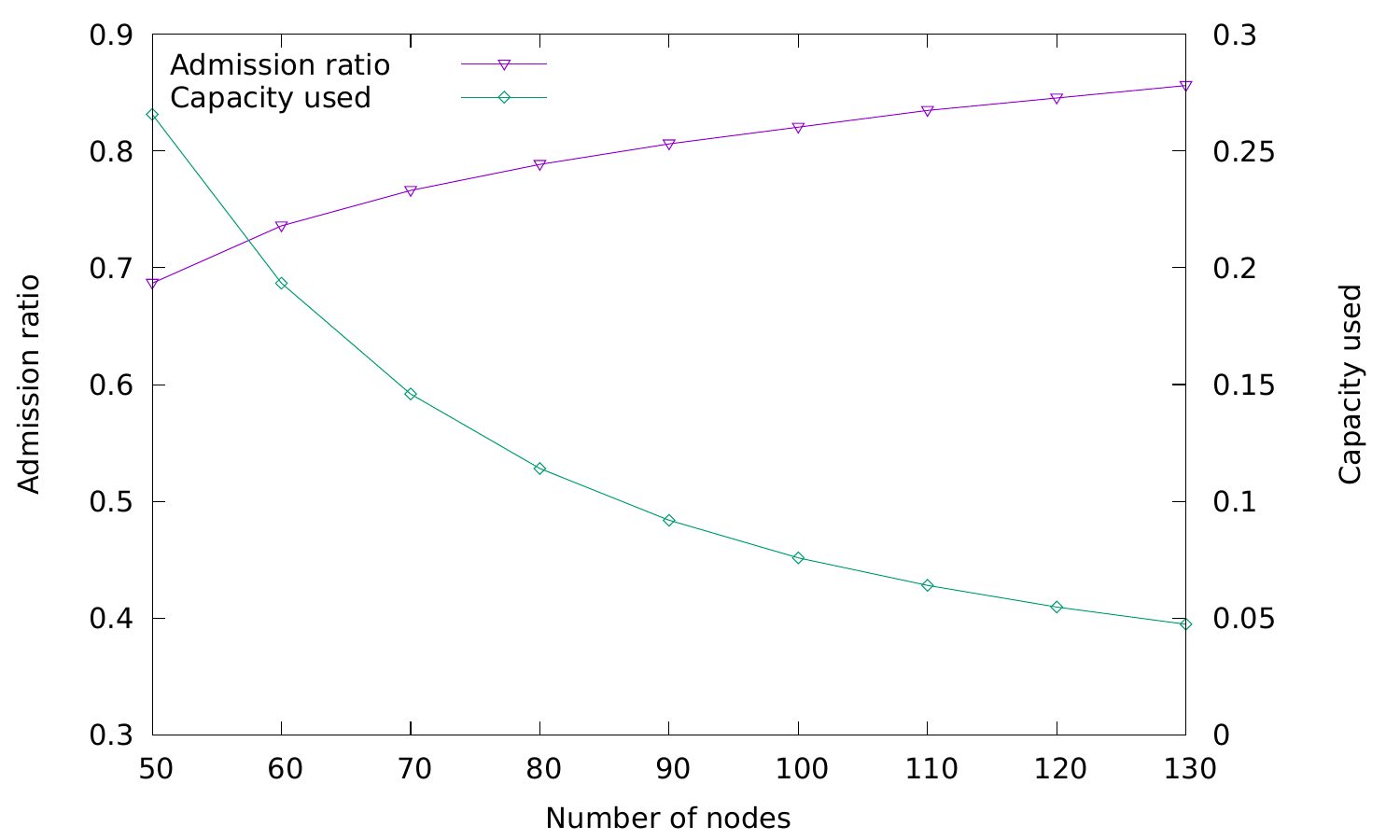}{Admission ratio and capacity used with increasing number of nodes, with \texttt{SPF} policy, $\alpha=0.4$, and 500 applications.}

Finally, we perform a sensitivity analysis on the network topology parameters.
In the following we only show the results with \texttt{SPF}.
In \rfig{006-va-admission} we show the admission ratio and capacity used when varying $\alpha$, while keeping all other parameters constant: 50 nodes and 300 applications, which is a high-load condition.
With increasing $\alpha$ we have a higher density of shorter edges, which explains why the admission ratio increases.
However, this comes at the cost of an increasingly large fraction of the QKD network capacity remaining unused.
In \rfig{006-vn-admission} we show the same metrics when increasing the network size, with $\alpha = 0.4$ and 500 applications, i.e., moderate load.
Also in this case we see that a small increase of the admission ratio requires substantial sacrifices in terms of unused network capacity.
\keypoint{There is a trade-off between the network resources (density, nodes) and its efficiency, in terms of capacity used/available.}
  \section{Conclusions}%
  \label{sec:conclusions}%
  In this paper we have raised awareness about the emergence of QKD networks, which allow the secure exchange of secret material to encrypt communications, and are close to achieving mass production, at least with trusted relay nodes.
We have investigated the use case of edge computing infrastructures and defined mathematical models for the resource allocation, which have led to the formulation of a practical problem for the admission control of incoming edge applications.
We have proposed three online policies that depend on the relations between the network operator and service provider in the real system.
We have evaluated the solutions using Monte Carlo simulations with unstructured network topologies and a trace-driven workload composition.
The results have shown that an interplay exists between resource allocation by the network operator and service provider, which should be exploited to attain high utilization and low rejection rate.

\optional{The work presented can be extended in many directions, including: broader analysis of the online policies defined to assess the performance with different models for the topology and workload composition; definition of management interfaces for the implementation in production enviroments; extension of the model to the case of serverless computing; integration with L2/L3 QKD routing protocols; generalization to untrusted relay nodes.
Furthermore, we are working on the realization of a testbed with a hybrid emulated/real QKD infrastructure for the performance evaluation in more realistic conditions.}%

\optional{
\section*{Acknowledgment}

The work of C.~Cicconetti was partially supported by project SERICS (PE00000014) under the MUR National Recovery and Resilience Plan funded by the European Union -- NextGenerationEU.
The work of M.~Conti and A.Passarella was co-funded by EU, \textit{PON Ricerca e Innovazione} 2014--2020 FESR/FSC Project ARS01\_00734 QUANCOM.
}


\begin{acronym}
  \acro{3GPP}{Third Generation Partnership Project}
  \acro{5G-PPP}{5G Public Private Partnership}
  \acro{AA}{Authentication and Authorization}
  \acro{ADF}{Azure Durable Function}
  \acro{AI}{Artificial Intelligence}
  \acro{API}{Application Programming Interface}
  \acro{AP}{Access Point}
  \acro{AR}{Augmented Reality}
  \acro{BGP}{Border Gateway Protocol}
  \acro{BSP}{Bulk Synchronous Parallel}
  \acro{BS}{Base Station}
  \acro{CDF}{Cumulative Distribution Function}
  \acro{CFS}{Customer Facing Service}
  \acro{CPU}{Central Processing Unit}
  \acro{DAG}{Directed Acyclic Graph}
  \acro{DHT}{Distributed Hash Table}
  \acro{DNS}{Domain Name System}
  \acro{DRR}{Deficit Round Robin} 
  \acro{ETSI}{European Telecommunications Standards Institute}
  \acro{FCFS}{First Come First Serve}
  \acro{FSM}{Finite State Machine}
  \acro{FaaS}{Function as a Service}
  \acro{GPU}{Graphics Processing Unit}
  \acro{HTML}{HyperText Markup Language}
  \acro{HTTP}{Hyper-Text Transfer Protocol}
  \acro{ICN}{Information-Centric Networking}
  \acro{IETF}{Internet Engineering Task Force}
  \acro{IIoT}{Industrial Internet of Things}
  \acro{ILP}{Integer Linear Programming}
  \acro{IPP}{Interrupted Poisson Process}
  \acro{IP}{Internet Protocol}
  \acro{ISG}{Industry Specification Group}
  \acro{ITS}{Intelligent Transportation System}
  \acro{ITU}{International Telecommunication Union}
  \acro{IT}{Information Technology}
  \acro{IaaS}{Infrastructure as a Service}
  \acro{IoT}{Internet of Things}
  \acro{JSON}{JavaScript Object Notation}
  \acro{KMS}{Key Management System}
  \acro{K8s}{Kubernetes}
  \acro{KPI}{Key Performance Indicator}
  \acro{KVS}{Key-Value Store}
  \acro{LCM}{Life Cycle Management}
  \acro{LL}{Link Layer}
  \acro{LOCC}{Local Operations and Classical Communication}
  \acro{LP}{Linear Programming}
  \acro{LTE}{Long Term Evolution}
  \acro{MAC}{Medium Access Layer}
  \acro{MANET}{Mobile Ad-Hoc Network}
  \acro{MBWA}{Mobile Broadband Wireless Access}
  \acro{MCC}{Mobile Cloud Computing}
  \acro{MDI}{Measurement-Device independent}
  \acro{MEC}{Multi-access Edge Computing}
  \acro{MEH}{Mobile Edge Host}
  \acro{MEPM}{Mobile Edge Platform Manager}
  \acro{MEP}{Mobile Edge Platform}
  \acro{ME}{Mobile Edge}
  \acro{ML}{Machine Learning}
  \acro{MNO}{Mobile Network Operator}
  \acro{NAT}{Network Address Translation}
  \acro{NISQ}{Noisy Intermediate-Scale Quantum}
  \acro{NFV}{Network Function Virtualization}
  \acro{NFaaS}{Named Function as a Service}
  \acro{OSPF}{Open Shortest Path First}
  \acro{OSS}{Operations Support System}
  \acro{OS}{Operating System}
  \acro{OWC}{OpenWhisk Controller}
  \acro{PMF}{Probability Mass Function}
  \acro{PPP}{Poisson Point Process}
  \acro{PU}{Processing Unit}
  \acro{PaaS}{Platform as a Service}
  \acro{PoA}{Point of Attachment}
  \acro{PPP}{Poisson Point Process}
  \acro{QC}{Quantum Computing}
  \acro{QKD}{Quantum Key Distribution}
  \acro{QoE}{Quality of Experience}
  \acro{QoS}{Quality of Service}
  \acro{QR}{Quantum Repeater}
  \acro{QWAP}{Quantum Workers' Assignment Problem}
  \acro{RN}{Relay Node}
  \acro{RPC}{Remote Procedure Call}
  \acro{RR}{Round Robin}
  \acro{RSU}{Road Side Unit}
  \acro{SBC}{Single-Board Computer}
  \acro{SDK}{Software Development Kit}
  \acro{SDN}{Software Defined Networking}
  \acro{SJF}{Shortest Job First}
  \acro{SLA}{Service Level Agreement}
  \acro{SMP}{Symmetric Multiprocessing}
  \acro{SoC}{System on Chip}
  \acro{SLA}{Service Level Agreement}
  \acro{SRPT}{Shortest Remaining Processing Time}
  \acro{SPT}{Shortest Processing Time}
  \acro{STL}{Standard Template Library}
  \acro{SaaS}{Software as a Service}
  \acro{TCP}{Transmission Control Protocol}
  \acro{TSN}{Time-Sensitive Networking}
  \acro{UDP}{User Datagram Protocol}
  \acro{UE}{User Equipment}
  \acro{URI}{Uniform Resource Identifier}
  \acro{URL}{Uniform Resource Locator}
  \acro{UT}{User Terminal}
  \acro{VANET}{Vehicular Ad-hoc Network}
  \acro{VIM}{Virtual Infrastructure Manager}
  \acro{VR}{Virtual Reality}
  \acro{VM}{Virtual Machine}
  \acro{VNF}{Virtual Network Function}
  \acro{WLAN}{Wireless Local Area Network}
  \acro{WMN}{Wireless Mesh Network}
  \acro{WRR}{Weighted Round Robin}
  \acro{YAML}{YAML Ain't Markup Language}
\end{acronym}




\end{document}